\documentclass[conference]{IEEEtran}
\IEEEoverridecommandlockouts
\usepackage{cite}
\usepackage{multirow}
\usepackage{booktabs} % Required for \toprule, \midrule, \bottomrule
\usepackage[table]{xcolor}   % provides \cellcolor
\usepackage{hyperref}
\definecolor{heatgreen}{HTML}{63BE7B}   % gradient endpoint (best value)
\definecolor{utmoscream}{HTML}{FCEFCB}  % empty O-UTMOS placeholder
\usepackage{amsmath,amssymb,amsfonts}
\usepackage{algorithmic}
\usepackage{graphicx}
\usepackage{textcomp}
\usepackage{xcolor}
\usepackage[most]{tcolorbox}
\usepackage{url}
\def\BibTeX{{\rm B\kern-.05em{\sc i\kern-.025em b}\kern-.08em
    T\kern-.1667em\lower.7ex\hbox{E}\kern-.125emX}}
    
\begin{document}
% \title{Singlish TTS: Training and Evaluation}
\title{Singlish, Can or Not? Fine-Tuning and Evaluating Zero-Shot TTS for Singapore English}

% \author{\IEEEauthorblockN{Ivan Kukanov}
% \IEEEauthorblockA{%\textit{dept. name of organization (of Aff.)} \\
% \textit{KLASS Engineering and Solutions}\\
% Singapore \\
% ivan.kukanov@klasses.com.sg}
% \and
% \IEEEauthorblockN{Zheng Xin Chai}
% \IEEEauthorblockA{%\textit{dept. name of organization (of Aff.)} \\
% \textit{KLASS Engineering and Solutions}\\
% Singapore \\
% email address or ORCID}
% }

% For camera ready
\author{
\IEEEauthorblockN{
Ivan Kukanov, Zheng Xin Chai}
\IEEEauthorblockA{
KLASS Engineering \& Solutions, Singapore \quad}
\IEEEauthorblockA{
\{ivan.kukanov, zhengxin.chai\}@klasses.com.sg
}
}

\graphicspath{{imgs}}

\maketitle
\bstctlcite{IEEEexample:BSTcontrol}

% Possible Titles
% Teaching Zero-Shot TTS to Speak Singlish: Fine-Tuning and Evaluation

% From Single-Shot to Fine-Tuning: Closing the Singlish Accent Gap in Zero-Shot TTS

% Can Zero-Shot TTS Clone an Accent? Fine-Tuning Chatterbox and CosyVoice 3 for Singlish

% Off-the-Shelf TTS Cannot Speak Singlish: Quantifying the Effect of Fine-Tuning on Accent Fidelity

% \textbf{Singlish, Can or Not? Fine-Tuning and Evaluating Zero-Shot TTS for Singapore English}

% Fine-Tuning Zero-Shot TTS for Singapore English: A Multi-Metric Evaluation of Accent Adaptation

% Q: why CV reduced on SPK-SIM vs F-CV SPK-SIM?
% Q: why WER for CV is high for off-shelf
% TODO: try wav2vec for O-ACC-SIM

\begin{abstract}
Zero-shot text-to-speech (ZS-TTS) achieves near-human quality for standard English, but it copies regional accents poorly. Prompted with a short Singlish utterance, state-of-the-art systems reproduce a speaker's timbre while flattening the accent toward generic English. We investigate whether targeted fine-tuning off-the-shelf ZS-TTS can close the gap for Singapore English (Singlish). We fine-tune two cutting-edge ZS-TTS models, Chatterbox and CosyVoice~3, on 50 Singlish speakers from the Infocomm Media Development Authority National Speech Corpus (IMDA NSC). Three speech distributions are evaluated: real recordings against off-the-shelf and fine-tuned generation driven by the same Singlish audio prompts. The evaluation covers four dimensions: naturalness, intelligibility, speaker similarity, and accent similarity. We separate adaptation (in-domain speakers seen during fine-tuning) from consistency (held-out speakers) to test whether accent transfer generalises beyond the training data. Fine-tuning raises accent similarity on in-domain and out-of-domain speakers for both Chatterbox and CosyVoice~3. It moves the generated distribution measurably toward real Singlish, with the gain persisting on held-out speakers. To our knowledge, this is the first systematic study of Singlish-accented TTS.

% [Results: Fine-tuning raises accent  similarity by 25\% on in-domain and 20\% on out-of-domain speakers for Chatterbox. It moves the generated distribution measurably toward real Singlish, with the gain persisting on held-out speakers. TODO: numbers] To our knowledge, this is the first systematic study of Singlish-accented TTS.
\end{abstract}


\section{Introduction}

Zero-shot text-to-speech (ZS-TTS) has matured rapidly. Large codec-language and flow-matching models - \mbox{VALL-E}~\cite{chen2025neural,chen2024valle2}, the CosyVoice family~\cite{Du2024CosyVoiceAS,Du2024CosyVoice2S,Du2025CosyVoice3T}, \mbox{F5-TTS}~\cite{chen-etal-2025-f5}, NaturalSpeech~3~\cite{ju2024naturalspeech3}, and the recent open release of Chatterbox~\cite{chatterboxtts2025} - can now clone a speaker from a few seconds of reference audio and synthesize natural-sounding speech in dozens of languages, with several systems claiming human parity on native-English benchmarks. This progress, however, is unevenly distributed across the world's English varieties. Recent works show that ZS-TTS still falls short in accent fidelity for non-native and regional variants~\cite{Zhong2025} \cite{poon2025clarity}, and that timbre, prosody, and accent remain entangled in the model's latent representations~\cite{ju2024naturalspeech3}. Inclusive TTS for under-represented accent communities is an open scientific problem and a pressing practical need. 

\begin{figure}[!htb]
    \centerline{\hspace{-0.2cm}\includegraphics[trim={0 0.9cm 0 0.9cm}, clip, scale=0.65]{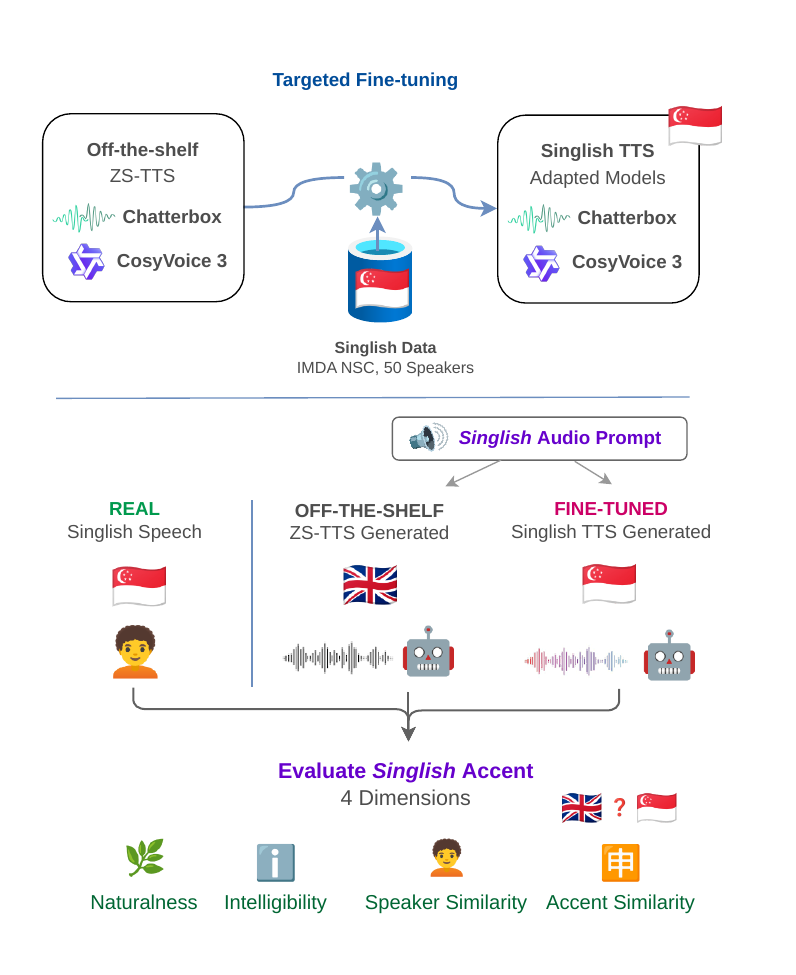}}
    \caption{Off-the-shelf zero-shot text-to-speech models (Chatterbox, CosyVoice~3) fail to transfer the Singlish accent when provided with an audio prompt. After fine-tuning on a Singlish dataset, we measure whether this gap is closed by evaluating four dimensions: naturalness, intelligibility, speaker and accent similarity.}
    \label{fig:graphical_abs}
\end{figure}

Singapore English (or colloquial register Singlish) is an example of an accent that today's TTS systems struggle to reproduce. Shaped by sustained contact with Mandarin, Malay, and Tamil, Singlish exhibits well-documented phonological differences from British or American English,
% monophthongisation of the FACE and GOAT vowels, conflated vowel-length pairs, dental to alveolar stop substitutions, glottalisation of final stops, and a strongly syllable-timed rhythm , 
together with a rich inventory of discourse particles (\textit{lah}, \textit{leh}, \textit{lor}, \textit{meh})~\cite{deterding2007singapore,kalaivanan2021homogenization}. The Singapore speech community has contributed heavily in Automatic Speech Recognition (ASR) and Audio Large Language Model (Audio-LLM) coverage of this variety, through the Infocomm Media Development Authority National Speech Corpus (IMDA NSC)~\cite{koh2019imda}, its multitask extension MNSC~\cite{wang2025mnsc}, MERaLiON-AudioLLM~\cite{he2025meralion}, and Singapore-specific speech encoders~\cite{meralion2024speechencoder}. The TTS side, however, has not received comparable attention. In our own experiments, off-the-shelf models prompted with a single Singlish reference produce English that is intelligible but non-Singlish: discourse particles are preserved at the lexical level, yet the prosody and segmental detail of the reference are not reliably transferred. To our knowledge, no peer-reviewed work has quantified this failure mode, nor studied what it takes to close the gap.

In this work, we systematically study zero-shot TTS targeted fine-tuning and evaluation for Singlish accent. We fine-tune two recent available LLM-style TTS systems Chatterbox~\cite{chatterboxtts2025} and CosyVoice 3~\cite{Du2025CosyVoice3T} on a pool of 50 Singlish speakers drawn from Part 3 of the IMDA NSC~\cite{koh2019imda}.
%, with approximately 40 minutes of speech per speaker after ASR-based WER filtering with a Singlish-fine-tuned Whisper [Wong, 2024]. 
We then compare three distributions: (i) real Singlish reference recordings, (ii) speech generated by the fine-tuned models, and (iii) speech generated by the same models off-the-shelf with identical Singlish audio prompts. Evaluation is split in two regimes. An \emph{adaptation regime}, in which the fine-tuned model is tested on the 50 in-domain speakers that was exposed to during fine-tuning; and a \emph{consistency regime}, in which it is tested on 42 held-out out-of-domain Singlish speakers that has never been seen. This split let us separate the question \textit{``did the model learn to imitate these particular speakers?''} from \textit{``did it learn the Singlish accent itself?''}

For evaluating the Singlish accent, we adopt the recently proposed \textit{CodecMOS-Accent} protocol~\cite{huang2026codecmos}, which validates four objective metrics: UT-MOS for naturalness, Word Error Rate (WER) for intelligibility, SPK-SIM for speaker similarity, and ACC-SIM for accent similarity. 
Our contributions are as follows:
\begin{itemize}
    \item \textbf{Singlish TTS adaptation and objective evaluation.}
    We provide the first quantitative characterisation of off-the-shelf ZS-TTS behaviour on Singlish and evaluate accent adaptation by fine-tuning Chatterbox and CosyVoice 3 on 50 Singlish speakers. We assess naturalness, intelligibility, speaker similarity, and accent similarity using four objective metrics: UT-MOS, WER, SPK-SIM, and \mbox{ACC-SIM}.

    \item \textbf{Adaptation vs. consistency evaluation.}
    We evaluate adaptation on 50 in-domain speakers seen during fine-tuning and consistency on 42 unseen out-of-domain speakers to assess whether accent adaptation generalizes beyond the training speakers.

    \item \textbf{Reproducible Singlish-TTS pipeline.}
    We describe a preprocessing pipeline for IMDA~NSC~Part~3~\cite{koh2019imda} corpus, including voice activity detection, ASR-based WER filtering using Singlish Whisper~\cite{wong2025whisper}, gender balancing, and signal-to-noise ratio (SNR) / speech quality and intelligibility measures (SQUIM) audio-quality screening.

    \item \textbf{Community contribution.}
    Both adapted Singlish TTS systems contributed to data generation for the RADAR~2026~\cite{luong2026radar} audio-deepfake detection challenge.
\end{itemize}

\begin{figure}[!htb]
\centerline{\hspace{-0.2cm}\includegraphics[trim={0 0.9cm 0 0.9cm}, clip, scale=0.6]{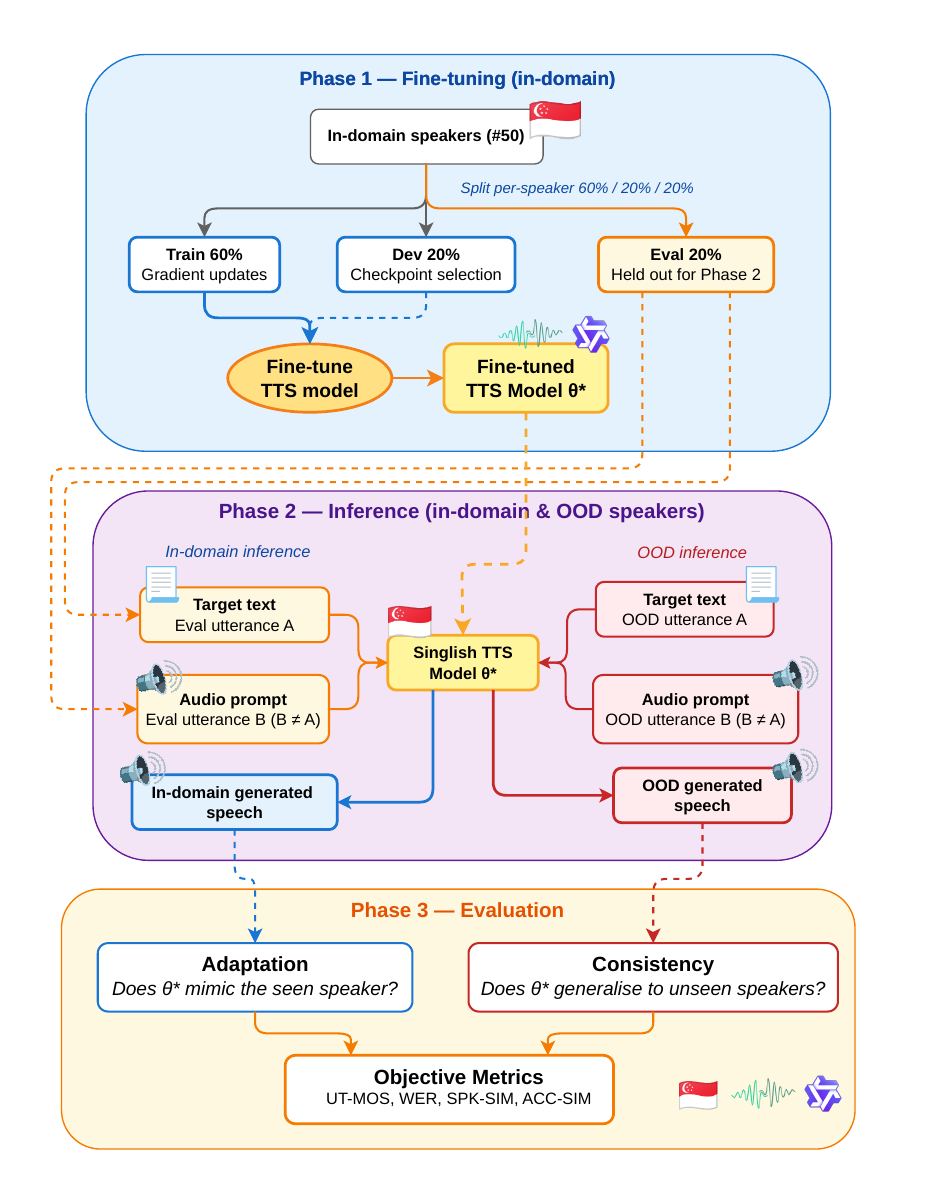}}
\caption{Three-phase experimental pipeline: fine-tuning TTS models on 50 in-domain Singlish speakers with per-speaker 60\%/20\%/20\% splits (Phase~1); inference on in-domain eval split utterances and unseen out-of-domain speakers, with audio prompt B disjoint from target A (Phase~2); and objective metrics evaluation of \textit{adaptation} to seen speakers and \textit{consistency} on unseen speakers (Phase~3).}
\label{fig:tts_flow}
\end{figure}

\section{Related Work}

\noindent\textbf{Zero-shot TTS and accent adaptation.}
% Modern zero-shot TTS clones a speaker from seconds of audio using codec-language or flow-matching backbones~\cite{chen2025neural,Du2025CosyVoice3T,chen-etal-2025-f5}, but accent fidelity lags behind speaker similarity because accent is entangled with timbre in the learned latents~\cite{ju2024naturalspeech3,Zhong2025}. Adapting such models to a target accent has been approached with lightweight parameter-efficient fine-tuning - adapters trained on a small fraction of weights, optionally with a distribution-matching loss~\cite{yang2023peltts} - and with explicit accent-embedding conditioning, as in AccentBox~\cite{Zhong2025}. YourTTS further showed that under a minute of target speech suffices to fine-tune a multi-speaker model to a new voice~\cite{casanova2022yourtts}. Closest to our setting, CLARITY targets accent bias in instruction-guided TTS across twelve English varieties, and includes a Singaporean-accented (SG) condition - drawn from the English-dominant portion of the SEAME code-switching corpus - on which it reports high perceptual quality and no significant bias~\cite{poon2025clarity}. Crucially, that SG condition is \emph{accented standard English}, not the colloquial Singlish register - with its characteristic discourse particles and syllable-timed prosody - that we target here. These efforts address either Mandarin-accented or broad L2 English, or the Singaporean accent in isolation; none models Singlish itself, and most evaluate adaptation only on seen speakers rather than separating it from cross-speaker generalisation.
Modern zero-shot TTS clones a speaker from seconds of audio using codec-language or flow-matching backbones~\cite{chen2025neural,Du2025CosyVoice3T,chen-etal-2025-f5}, but accent fidelity lags behind speaker similarity because accent is entangled with timbre in the learned latents~\cite{ju2024naturalspeech3,Zhong2025}. Prior approaches to accent adaptation fall into three groups. (i)~\emph{Adaptation on accented speech}: fine-tuning on multi-accent corpora with explicit accent conditioning via accent IDs, tokens, or reference audio, making accents selectable at inference~\cite{zhou2024accented,lux2022low,xinyuan2025scalable,Zhong2025}, or training lightweight adapters on a small fraction of weights~\cite{yang2023peltts}. (ii)~\emph{Adaptation without accented speech}: in~\cite{lertpetchpun2026accent} learn accent vectors from the speakers' native (L1) speech and combine them with the multilingual XTTS~2~\cite{casanova2024xtts} to synthesize accented English; the vectors can be scaled or interpolated to control accent strength. (iii)~\emph{Linguistic input transformation}: modifying the input text via phonological rules~\cite{lertpetchpun2026learning} or LLM-based transliteration into the accent's native writing system~\cite{inoue2025macst}, leaving the model untouched. 

\noindent\textbf{Singlish and Resources.}
Singapore English is diglossic~\cite{goh2016anatomy,harada2009roles}: formal Singapore Standard English (SSE) coexists with colloquial Singlish, which departs from SSE in simplified grammar, distinctive pronunciation, stress and intonation patterns, loanwords from Malay, Chinese, and Tamil, and pragmatic particles such as \textit{lah}, \textit{lor}, and \textit{leh}~\cite{ningsih2023exploring,kareba2022singlish,peng2013singlish}. Singlish has been described phonologically~\cite{deterding2007singapore,kalaivanan2021homogenization} and is supported by the IMDA NSC~\cite{koh2019imda}, its multitask annotation MNSC~\cite{wang2025mnsc}, and Singapore-tuned models such as MERaLiON-AudioLLM~\cite{he2025meralion} and a dedicated speech encoder~\cite{meralion2024speechencoder}. This activity concentrates on recognition and understanding; synthesis of Singlish is, to our knowledge, unaddressed.

\noindent\textbf{Evaluation methods.}
Speaker similarity is conventionally measured as cosine similarity in an ECAPA-TDNN embedding space~\cite{desplanques2020ecapa}; the same encoder family fine-tuned on CommonAccent~\cite{zuluaga2023commonaccent} has been shown to encode \emph{accent}, motivating our accent-similarity score. Naturalness is commonly proxied by UT-MOS~\cite{saeki2022utmos}. Beyond per-utterance scores, TTSDS and TTSDS2 compare real and synthetic speech as distributions over prosody, speaker, and intelligibility factors~\cite{minixhofer2024ttsds,minixhofer2025ttsds2robust}. \textit{CodecMOS-Accent} recently unified intelligibility, speaker-, accent-, and naturalness measures into a single objective protocol validated against subjective ratings on accented English~\cite{huang2026codecmos}. We adopt this protocol for benchmarking Singlish for the first time.

\section{Singlish Dataset}\label{sec:sg_data}

We derive Singlish speech dataset from Part~3 of IMDA~NSC~\cite{koh2019imda} Speech-to-Text, which consists of $\sim$1000 hours of conversational data recorded from $\sim$500 pairs of local English speakers. The conversations 
cover everyday topics and game-playing activities. For this study, we use only close-talk microphone recordings from the Same Room subset in which both speakers were recorded in the same room. Specifically, we used the Huggingface mesolitica/IMDA-STT~\footnote{\href{https://huggingface.co/datasets/mesolitica/IMDA-STT}{https://huggingface.co/datasets/mesolitica/IMDA-STT}} dataset version. Each utterance is provided as a separate MP3 file, with its corresponding transcript in the metadata.

We then checked the quality of the samples by performing Singlish Whisper ASR~\footnote{\href{https://huggingface.co/mjwong/whisper-large-v3-turbo-singlish}{https://huggingface.co/mjwong/whisper-large-v3-turbo-singlish}} on the utterances and then computing the WER between each ground-truth transcript and the corresponding ASR output. Afterward, we excluded samples with WER greater than 50\%, as manual inspection indicated that their utterances often did not align well with the ground-truth transcripts.

\begin{figure}[htbp]
    \centerline{\hspace{-0.2cm}\includegraphics[scale=0.4]{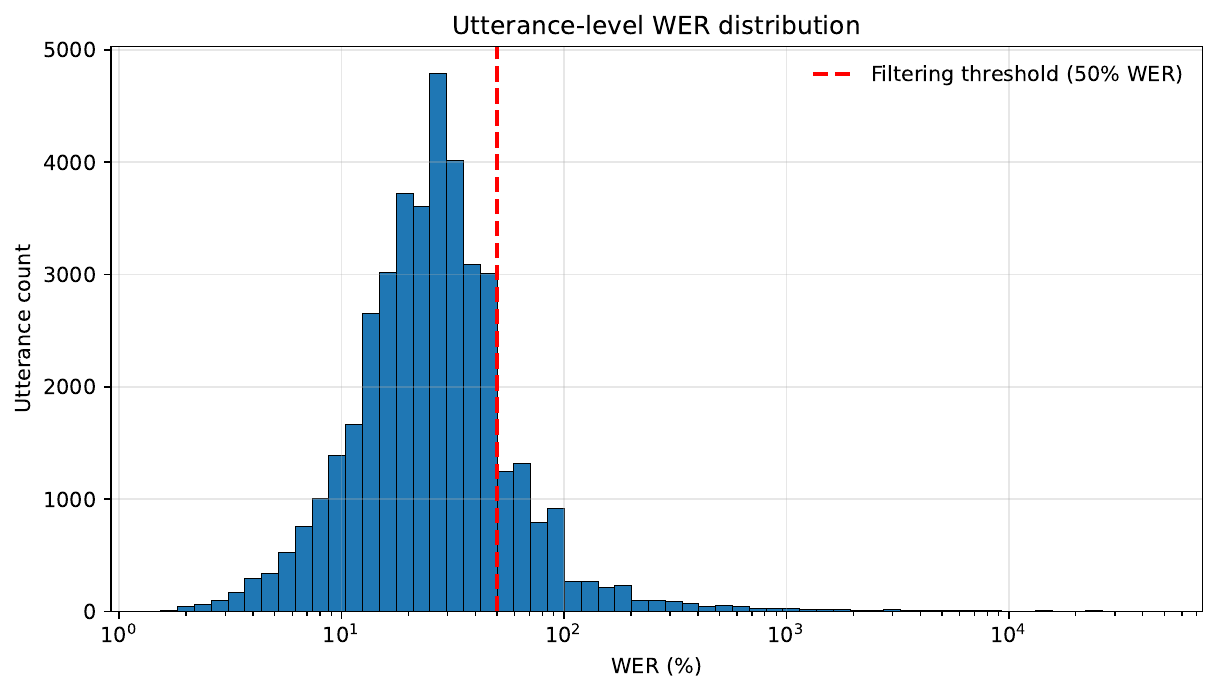}}
    \caption{Utterance-based WER (\%) (no text normalization) histogram of Singlish Whisper ASR output on IMDA NSC Part~3 Same Room subset~\cite{koh2019imda}. After thresholding at 50\% WER, 85.7\% of utterances are retained.}
    \label{fig:placeholder}
\end{figure}

\begin{figure}[htbp]
    \centerline{\hspace{-0.2cm}\includegraphics[width=\linewidth]{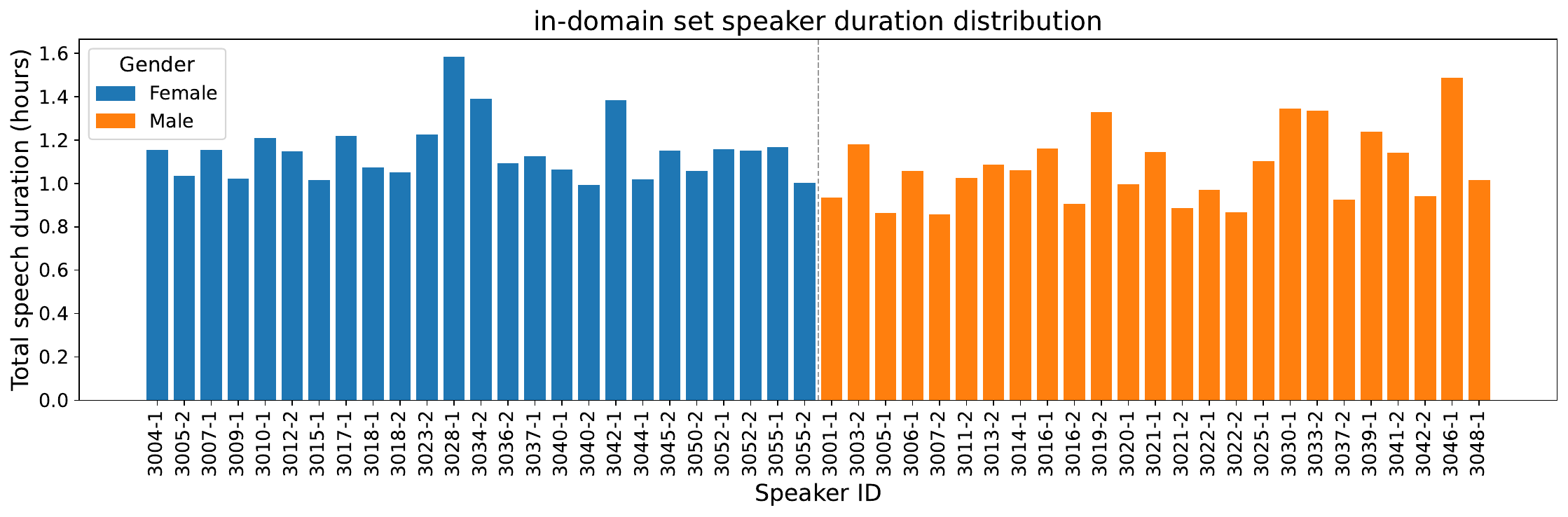}}
    \centerline{\hspace{-0.2cm}\includegraphics[width=\linewidth]{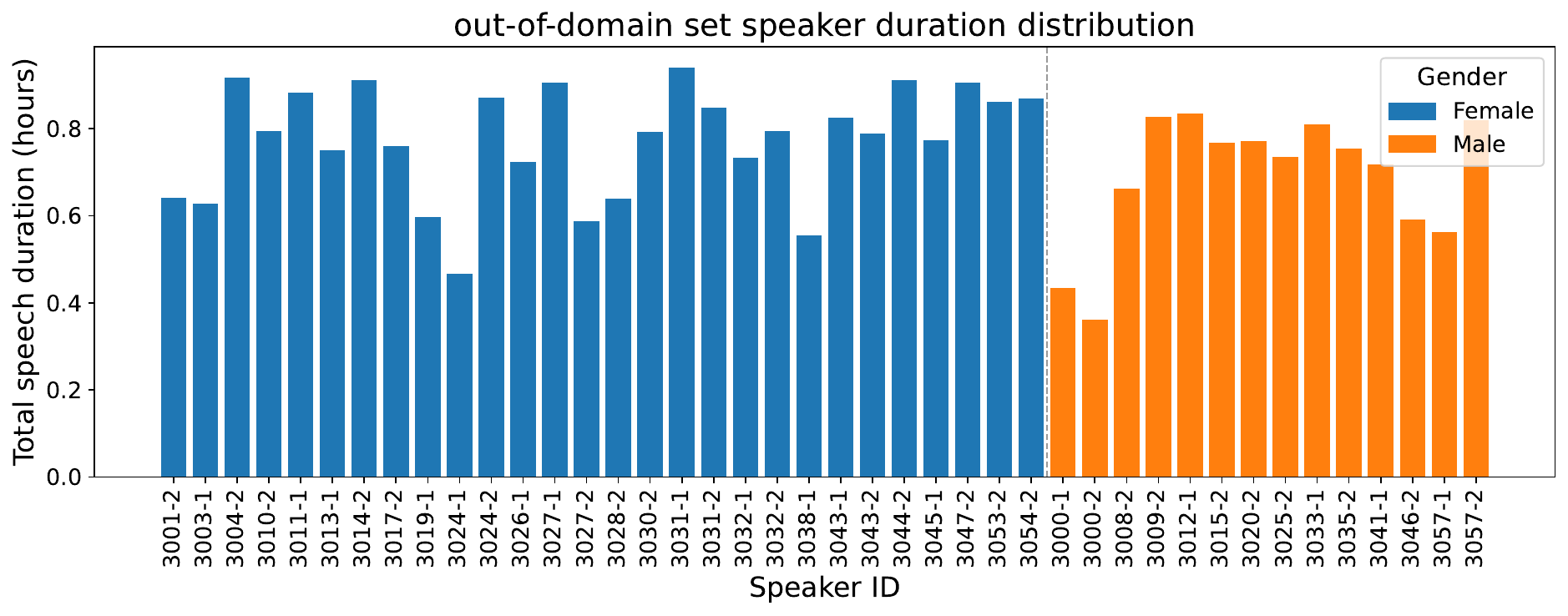}}    
    \caption{Speaker duration distribution of in-domain and out-of-domain sets.}
    \label{fig:speaker_dur_dist}
\end{figure}

Lastly, we selected 50 speakers to form our in-domain set and another 42 speakers to form our out-of-domain set, Fig.~\ref{fig:speaker_dur_dist}. All utterances of a selected speaker are placed in the corresponding set, guaranteeing no speaker overlap. Fig.~\ref{fig:speaker_dur_dist} shows the speaker duration distributions of both sets. Male and female are kept balanced in in-domain while there are roughly twice as many female speakers as male in out-of-domain. The total duration is 56.80 hours and average speaker duration is 1.14 hours. More statistics of speakers see in Tab.~\ref{tab:spk_stats}.

% \begin{table}[htbp]
% \caption{Speaker duration(h) statistics for each set}
% \begin{center}
% \begin{tabular}{|c|c|c|c|c|c|c|}
%         \hline
%         \textbf{Set} & \textbf{Mean} & \textbf{Std} & \textbf{Min} & \textbf{Max} & \textbf{Sum} & \textbf{\# of speakers} \\
%         \hline
%         ID & 1.11 & 0.16 & 0.86 & 1.59 & 55.50 & 50 \\
%         \hline
%         OOD & 0.75 & 0.14 & 0.36 & 0.94 & 31.32 & 42 \\
%         \hline
% \end{tabular}
% \label{tab:spk_stats}
% \end{center}
% \end{table}

\begin{table}[htbp]
\caption{In-/out-of-domain speech statistics}
\begin{center}
\begin{tabular}{c c c c c c c }
        \toprule
        \textbf{Set} & \textbf{Mean, h} & \textbf{Total, h} & \textbf{\#Utt} & \textbf{\#Spk} & \textbf{\#M} & \textbf{\#F} \\
        \hline
        in-domain & 1.14 & 56.80 & 19,626 & 50 & 25 & 25 \\
        % \hline
        out-of-domain & 0.72 & 30.43 & 13,505 & 42 & 14 & 28 \\
        % \hline
        \bottomrule
\end{tabular}
\label{tab:spk_stats}
\end{center}
\end{table}

% \begin{figure}[htbp]
%     \centerline{\hspace{-0.2cm}\includegraphics[scale=0.3]{imgs/gender_distribution.pdf}}
%     \caption{Gender duration distribution of ID and OOD sets}
%     \label{fig:placeholder}
% \end{figure}

\section{Experimental Setup}

% \textbf{Experiments}
% 1. \textbf{Evaluate objective metrics}: compare real IMDA speech vs generated speech with off-the-shelf model vs generated speech with finetuned model. Then we will see how the capability of off-shelf model vs finetuned models to pickup the Singlish accent
% 2. Analyze outliers speech, abnormalities in distributions

\subsection{Dataset}

All experiments are based on the IMDA NSC~\cite{koh2019imda}, preprocessed in Sec. \ref{sec:sg_data}, which split into two disjoint groups: 50 \emph{in-domain} speakers and 42 \emph{out-of-domain} speakers who are entirely held out. \textit{In-domain} set uses a per-speaker $60\%/20\%/20\%$ split; the reference is taken from the evaluation split and removed from the evaluation targets, so the prompt utterance is never itself synthesized. Synthesis target transcripts come from evaluation split.
Both Chatterbox~\cite{chatterboxtts2025} and CosyVoice~3~\cite{Du2025CosyVoice3T} are fine-tuned on the identical in-domain pool: training split is for models finetuning and validation split is for hyper parameters selection. Therefore, any difference in adaptation behaviour is attributable to the models rather than the data. \textit{Out-of-domain} set speakers are entirely unseen during fine-tuning. Therefore, evaluation on the set tests whether the model has learned a consistent Singlish accent that generalizes across speakers.

At the \textit{inference} phase in Fig.~\ref{fig:tts_flow}, the generation is \textit{zero-shot voice cloning}: every utterance of a speaker is generated from a single reference recording of that speaker, with no speaker-specific adaptation at test time. The reference is chosen as the utterance whose duration is that speaker's mean duration, $\sim 8$ seconds duration. The audio prompts of every speaker are held-out from evaluation split of \textit{in-domain} subset or \textit{out-of-domain} subset.

% \begin{figure}[htbp]
% \centerline{\hspace{-0.2cm}\includegraphics[scale=0.4]{in_domain_utmos.png}}
% \caption{In-domain UT-MOS distribution.}
% \label{fig:id_utmos}
% \end{figure}

% \begin{figure}[htbp]
% \centerline{\hspace{-0.2cm}\includegraphics[scale=0.4]{ood_utmos.png}}
% \caption{Out-of-domain UT-MOS distribution.}
% \label{fig:ood_utmos}
% \end{figure}

\subsection{Fine-tuning CosyVoice~3}
Fine-tuning was implemented by modifying the official CosyVoice~3 training script and configuration provided by the authors \cite{cosyvoice_github_4d7295a}. Two fine-tuning variants were evaluated, both following the default CosyVoice~3 training configuration with some modifications. The first, \textit{F-CosyVoice-idx}, used the Speaker ID prompt together with speaker embeddings. The second, \textit{F-CosyVoice}, used the Default prompt with speaker embeddings disabled to better match the zero-shot inference setting used during evaluation. Unless otherwise stated, all training settings followed the default configuration.

The modified settings are summarized in Tab.~\ref{tab:cosyvoice_finetuning_params}. The two prompt variants are shown below.

\begin{tcolorbox}[title=Speaker ID prompt,
                  colback=gray!5,
                  colframe=gray!60,
                  boxrule=0.5pt,
                  left=2pt,right=2pt,top=2pt,bottom=2pt]
\small\ttfamily
You are speaker \{speaker ID\}.<|endofprompt|>
\end{tcolorbox}

\begin{tcolorbox}[title=Default prompt,
                  colback=gray!5,
                  colframe=gray!60,
                  boxrule=0.5pt,
                  left=2pt,right=2pt,top=2pt,bottom=2pt]
\small\ttfamily
You are a helpful assistant.<|endofprompt|>
\end{tcolorbox}

We tuned CosyVoice~3's hyperparameter settings and the choice of trainable components by randomly sampling five validation examples from the in-domain validation set. Hyperparameters and trainable components were optimized progressively rather than through an exhaustive grid search. Specifically, we progressively tuned 1 component while keeping the other components frozen, starting tuning from the LLM component to finally the vocoder. The final configuration was chosen out of the candidates by the second author based on the naturalness of the generated speech and its similarity to the corresponding ground-truth recordings.

During this tuning process, we observed that fine-tuning the LLM produced the most noticeable improvement in accent adaptation. Fine-tuning the Flow Matching model produced no noticeable perceptual changes, whereas fine-tuning the vocoder consistently degraded speech quality and introduced static-like artifacts. Consequently, the final configuration fine-tuned both the LLM and Flow Matching model while keeping the vocoder frozen.

Based on the tuning procedure and observations described above, the final hyperparameter settings and choice of trainable components used for all subsequent experiments are summarized in Tab.~\ref{tab:cosyvoice_finetuning_params}.

Following the default CosyVoice~3 training pipeline, all reported CosyVoice~3 results use the checkpoint obtained by element-wise averaging the model weights from the five best validation epochs.

\begin{table}[htbp]
\caption{F-CosyVoice-idx and F-CosyVoice fine-tuning setup and hyperparameters}
\begin{center}
\begin{tabular}{lcc}
\toprule
\textbf{Parameter} & \textbf{F-CosyVoice-idx} & \textbf{F-CosyVoice} \\
\midrule
LLM & FT & FT \\
Flow Matching & FT & FT \\
Vocoder & Frozen & Frozen \\
LR & $5\times10^{-7}$ & $5\times10^{-7}$ \\
Max batch frames & 2000 & 2000 \\
Epochs & 200 & 200 \\
Optimizer & Adam & Adam \\
Scheduler & Constant & Constant \\
Warmup steps & 2500 & 2500 \\
Grad clip & 5 & 5 \\
Grad accumulation & 2 & 2 \\
Speaker embedding & True & False \\
LLM prompt & Speaker ID & Default \\
\bottomrule
\end{tabular}
\label{tab:cosyvoice_finetuning_params}
\end{center}
\end{table}

\subsection{Inferencing CosyVoice~3}
During the inference phase in Fig.~\ref{fig:tts_flow}, speech is synthesized by three CosyVoice~3 systems: the off-the-shelf zero-shot baseline, \textit{CosyVoice}, and the two Singlish fine-tuned variants, \textit{F-CosyVoice-idx} and \textit{F-CosyVoice}, in Tab.~\ref{tab:total_metrics}. 

\textit{CosyVoice} and \textit{F-CosyVoice} operate in zero-shot mode, which conditions speech generation on a prompt speech and its corresponding transcript. In contrast, \textit{F-CosyVoice-idx} operates in speaker fine-tuned (SFT) mode, which conditions speech generation solely on a speaker ID without requiring prompt speech. Since speaker IDs are only learned for speakers seen during fine-tuning, \textit{F-CosyVoice-idx} is evaluated only on the in-domain set.

These three systems differ only in LLM and Flow Matching model weights, LLM prompt, and inference mode, as summarized in Tab.~\ref{tab:cosyvoice_inference_params}. All other decoding parameters are kept unchanged from the official CosyVoice~3 inference script~\cite{cosyvoice_github_4d7295a}.

\begin{table}[htbp]
\caption{CosyVoice, F-CosyVoice-idx and F-CosyVoice inference setup}
\begin{center}
\begin{tabular}{lccc}
\hline
\multicolumn{1}{l}{\textbf{Parameter}} & \multicolumn{1}{l}{\textbf{CosyVoice}} & \textbf{F-CosyVoice-idx} & \textbf{F-CosyVoice} \\ \hline
LLM                                    & Pretrained                             & FT                       & FT                   \\
Flow Matching                          & Pretrained                             & FT                       & FT                   \\
LLM prompt                             & Default                                & Speaker ID               & Default              \\
Inference mode                         & Zero-shot                              & SFT                      & Zero-shot           \\
\bottomrule
\end{tabular}
\label{tab:cosyvoice_inference_params}
\end{center}
\end{table}

\subsection{Fine-tuning Chatterbox}
% training data, inputs, model architecture, training criteria, hyperparams, stopping criteria, other specifics
% \textbf{T3: paper https://arxiv.org/html/2605.30748v2}
% ============
We adapt Chatterbox~\cite{chatterboxtts2025} TTS to Singlish by fine-tuning only its T3~\cite{seo2026chatterboxflashpriorcalibratedblockdiffusion} module - the autoregressive Llama-style~\cite{seo2026chatterboxflashpriorcalibratedblockdiffusion} transformer that maps text to discrete speech tokens while keeping the flow-matching decoder and the voice encoder frozen. For each utterance the text tokens are precomputed, $\sim$25Hz S3 speech tokens, an utterance-level speaker embedding, and a 3 seconds reference-prompt token sequence. The text vocabulary is extended to 2454 tokens. T3 module is optimized by next-token prediction with a combined cross-entropy over text and speech tokens. Detailed optimization hyper parameters were used from Tab.~\ref{tab:chatterbox_params}.
% We use AdamW (LR = $10^{-5}$, linear decay, no warmup), an effective batch size of 16, bf16 precision and gradient checkpointing, training for 100 epochs (73.6K steps). 
The training loss decreases from 6.0 to below $10^{-3}$ and the adequacy of the adapted model was confirmed by subjective listening test to a few samples synthesized from \textit{in-domain} validation split.

\begin{table}[htbp]
\caption{Chatterbox and F-Chatterbox Fine-Tuning Setup and Hyperparameters}
\begin{center}
\begin{tabular}{l p{0.66\columnwidth}}
\toprule
\textbf{Parameter} & \textbf{Value} \\
\midrule
% Data & IMDA NSC Part 3, 50 in-domain speakers, 11,775~utts \\
% Smp. Rate & 16KHz \\
% \hline
Tuned module & T3, full FT; S3Gen and VoiceEncoder frozen \\
% \hline
Vocab & Extended to 2,454 \\
% \hline
Loss & Text CE + Speech CE \\
\texttt{emotion\_adv} & 0.5 (emotion-exaggeration) \\
\hline
Optimizer & AdamW, LR=$10^{-5}$, Linear Decay, no warmup, bf16\\
% \hline
Batch Size & 16 \\
% \hline
Epochs & 100 epochs = 73,600 steps \\
% \hline
Loss curve & $6.0 \to {\sim}10^{-3}$ (after 100 epochs) \\
\bottomrule
\end{tabular}
\label{tab:chatterbox_params}
\end{center}
\end{table}

\begin{table*}[htbp]
\caption{Evaluation Results Across Systems and Domains}
\begin{center}
\begin{tabular}{c c c c c c c c }
\toprule
\multirow{2}{*}{\textbf{System}} & \multirow{2}{*}{\textbf{Subset}} & \multirow{2}{*}{\textbf{UT-MOS}} & \multirow{2}{*}{\textbf{WER, \%}} & \multicolumn{2}{c}{\textbf{SPK-SIM}} & \multicolumn{2}{c}{\textbf{ACC-SIM}} \\
\cline{5-8}
 & & & & \textbf{\textit{prompt}} & \textbf{\textit{match}} & \textbf{\textit{prompt}} & \textbf{\textit{match}} \\
\hline
\multirow{2}{*}{Ground Truth}
 & in-domain     & \cellcolor{heatgreen!14}2.50 & \cellcolor{heatgreen!33}17.31 & \cellcolor{heatgreen!31}0.7083 & \cellcolor{heatgreen!70}1.0000 & \cellcolor{heatgreen!45}0.6277 & \cellcolor{heatgreen!70}1.0000 \\
 & out-of-domain & \cellcolor{heatgreen!5}2.37 & \cellcolor{heatgreen!9}21.08 & \cellcolor{heatgreen!5}0.6543 & \cellcolor{heatgreen!70}1.0000 & \cellcolor{heatgreen!19}0.5802 & \cellcolor{heatgreen!70}1.0000 \\
\hline
\hline
\multirow{2}{*}{Chatterbox}
 & in-domain     & \cellcolor{heatgreen!69}3.34 & \cellcolor{heatgreen!70}\textbf{11.48} & \cellcolor{heatgreen!38}0.7238 & \cellcolor{heatgreen!14}0.6036 & \cellcolor{heatgreen!24}0.5890 & \cellcolor{heatgreen!10}0.5114 \\
 & out-of-domain & \cellcolor{heatgreen!69}3.33 & \cellcolor{heatgreen!56}13.73 & \cellcolor{heatgreen!21}0.6882 & \cellcolor{heatgreen!5}0.5364 & \cellcolor{heatgreen!5}0.5541 & \cellcolor{heatgreen!5}0.4697 \\
\hline
\multirow{2}{*}{CosyVoice}
 & in-domain     & \cellcolor{heatgreen!69}3.33 & \cellcolor{heatgreen!21}19.20 & \cellcolor{heatgreen!63}0.7774 & \cellcolor{heatgreen!22}0.6581 & \cellcolor{heatgreen!50}0.6368 & \cellcolor{heatgreen!18}0.5771 \\
 & out-of-domain & \cellcolor{heatgreen!70}\textbf{3.35} & \cellcolor{heatgreen!5}21.66 & \cellcolor{heatgreen!47}0.7425 & \cellcolor{heatgreen!12}0.5845 & \cellcolor{heatgreen!28}0.5959 & \cellcolor{heatgreen!11}0.5212 \\
\hline
\multirow{2}{*}{F-Chatterbox}
 & in-domain     & \cellcolor{heatgreen!30}2.75 & \cellcolor{heatgreen!45}15.39 & \cellcolor{heatgreen!70}\textbf{0.7913} & \cellcolor{heatgreen!27}0.6949 & \cellcolor{heatgreen!70}\textbf{0.6748} & \cellcolor{heatgreen!26}0.6376 \\
 & out-of-domain & \cellcolor{heatgreen!31}2.76 & \cellcolor{heatgreen!37}16.59 & \cellcolor{heatgreen!49}0.7463 & \cellcolor{heatgreen!15}0.6055 & \cellcolor{heatgreen!47}0.6324 & \cellcolor{heatgreen!16}0.5617 \\
\hline
\multirow{2}{*}{F-CosyVoice}
 & in-domain     & \cellcolor{heatgreen!63}3.24 & \cellcolor{heatgreen!50}14.68 & \cellcolor{heatgreen!54}0.7576 & \cellcolor{heatgreen!24}0.6739 & \cellcolor{heatgreen!41}0.6200 & \cellcolor{heatgreen!21}0.6036 \\
 & out-of-domain & \cellcolor{heatgreen!65}3.28 & \cellcolor{heatgreen!39}16.27 & \cellcolor{heatgreen!34}0.7148 & \cellcolor{heatgreen!13}0.5928 & \cellcolor{heatgreen!20}0.5820 & \cellcolor{heatgreen!15}0.5489 \\
\hline
F-CosyVoice-idx
 & in-domain     & \cellcolor{heatgreen!63}3.25 & \cellcolor{heatgreen!67}11.92 & \cellcolor{heatgreen!29}0.7043 & \cellcolor{heatgreen!33}\textbf{0.7383} & \cellcolor{heatgreen!20}0.5811 & \cellcolor{heatgreen!28}\textbf{0.6565} \\
\bottomrule
\end{tabular}
\label{tab:total_metrics}
\end{center}
\end{table*}

\begin{figure*}[!htbp]
\centerline{\hspace{-0.2cm}\includegraphics[trim={0 2.9cm 0 2.5cm}, clip, scale=0.2]{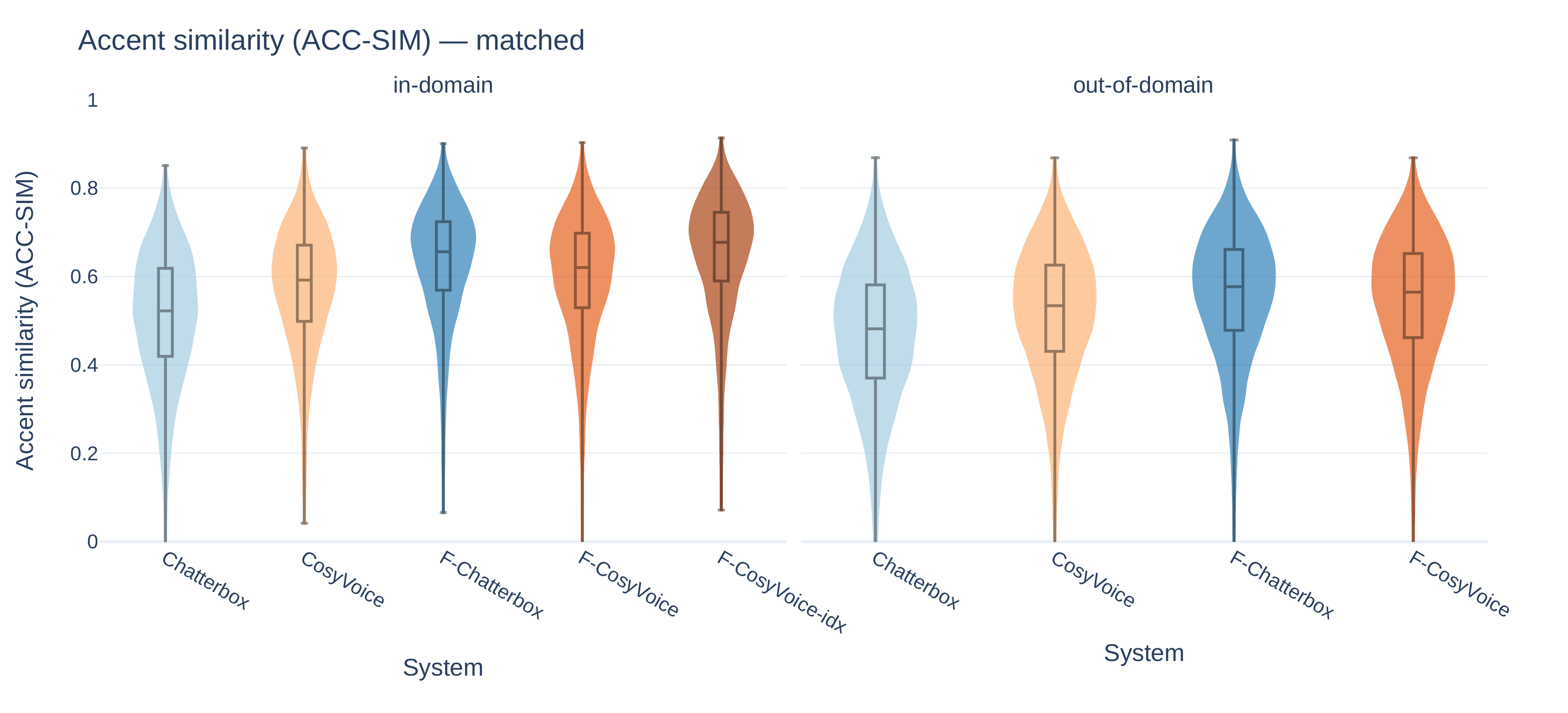}}
\caption{In-/out-of-domain  distribution across systems.}
\label{fig:violin_accent}
\end{figure*}

\subsection{Inference Chatterbox}

% - Inference workflow Fig. \ref{fig:tts_flow}: data splits, in-domain / out-of-domain, ZS-TTS / fine-tuned
% ========================
During the inference phase in Fig. \ref{fig:tts_flow}, speech is synthesized by two Chatterbox systems: the off-the-shelf zero-shot baseline, \textit{Chatterbox} and the Singlish fine-tuned model, \mbox{\textit{F-Chatterbox}}, in Tab.~\ref{tab:total_metrics}. These two systems share an identical pipeline and decoding configurations and differ only in the T3~\cite{seo2026chatterboxflashpriorcalibratedblockdiffusion} module weights, \mbox{\textit{F-Chatterbox}} uses fine-tuned T3. While S3Gen flow-matching and the VoiceEncoder modules remain the original pretrained modules in both.  
% The speaker embeddings are extracted using the VoiceEncoder; the target text is punctuation-normalized and tokenized. The T3 autoregressively generates S3 speech tokens with the \textit{temperature} of $0.75$ (the softmax sharpness of the speech-token sampler; lower is more deterministic generation). The classifier-free guidance weight, requiring the duplicated conditional/unconditional batch (logits = cond + $w$ * (cond - uncond)), \textit{cfg\_weight} is set to default of $0.5$. The \textit{repetition penalty} is $1.35$ and suppresses looping/stutter hallucinations. The \textit{exaggeration} of $0.4$ controls emotional intensity. 
The synthesizer hyper parameters, Tab.~\ref{tab:inf_chatterbox}, were selected on the in-domain validation split. 
% The result tokens are decoded by the frozen S3Gen flow-matching decoder and HiFi-GAN vocoder into a 24KHz waveform.

\begin{table}[htbp]
\centering
\caption{Chatterbox and F-Chatterbox inference hyperparameters}
\label{tab:hyperparams}
\begin{tabular}{lc}
\toprule
\textbf{Parameter} & \textbf{Value} \\
\midrule
\textit{temperature}         & 0.75 \\
\textit{cfg\_weight}         & 0.50 \\
\textit{repetition\_penalty} & 1.35 \\
\textit{exaggeration}        & 0.40 \\
\bottomrule
\end{tabular}
\label{tab:inf_chatterbox}
\end{table}
 
\subsection{Evaluation Metrics} 
% - Objective evaluation metrics\\
% - comparison real vs CB vs CV metrics Tab. \ref{tab:total_metrics}, Fig. \ref{fig:violin_accent} \\
% - strategies: prompt / matched
% ====================================
We evaluate real and generated datasets by all systems with the objective metrics \cite{huang2026codecmos} for Singlish accent capability benchmarking. We conduct objective evaluation using the following four commonly used metrics. Each similarity metric is reported in two forms. The \textit{prompt} variant is the cosine similarity between the synthesized utterance and the single-shot audio prompt (voice cloning reference) that conditioned generation. It measures how faithfully the system reproduces the specific reference voice. The \textit{match} variant is the cosine similarity of generated against the target ground-truth recording where generated and the real utterance have the same text content. It measures generalization rather than memorized prompt content. The Ground Truth scores exactly 1 on both \textit{match} columns in Tab.~\ref{tab:total_metrics}, since a real utterance is compared to itself and $<1$ on the prompt columns (cross-content, same-speaker reference baseline).

\textit{\textbf{Naturalness (UT-MOS)}}. We report naturalness using the UT-MOS neural predictor~\cite{saeki2022utmos}, an SSL-ensemble system from the VoiceMOS Challenge line of work \cite{cooper2023voicemos}. We treat \mbox{UT-MOS} cautiously: the predictor is trained predominantly on clean, read studio speech, which may limit its reliability when evaluating spontaneous Singlish recordings. UT-MOS is therefore read as a sanity check and a diagnostic of distance from clean-synthetic norms.
% , with TTSDS2 \cite{minixhofer2025ttsds2} providing the more trustworthy distributional view of how closely each system's output matches the real-speech reference across prosody, speaker, intelligibility, and general factors.

\textit{\textbf{Intelligibility (WER)}}. Word error rate is the standard intelligibility proxy across zero-shot TTS evaluation \cite{minixhofer2024ttsds,huang2026codecmos}. It measures whether generated Singlish speech remains decodable. We transcribe with the Singlish Whisper ASR model~\footnote{\href{https://huggingface.co/mjwong/whisper-large-v3-turbo-singlish}{https://huggingface.co/mjwong/whisper-large-v3-turbo-singlish}}.

\textit{\textbf{Speaker similarity (SPK-SIM)}}. It is the cosine similarity between ECAPA-TDNN speaker embeddings \cite{desplanques2020ecapa} of the synthesized and reference utterance. Where the reference is fixed reference audio prompt or all utterances of the ground truth dataset (prompt/match modes).

\textit{\textbf{Accent similarity (ACC-SIM)}}. Our headline metric is the cosine similarity in the accent-discriminative embedding space, computing similarity from the CommonAccent ECAPA-TDNN model\footnote{\href{https://huggingface.co/Jzuluaga/accent-id-commonaccent_ecapa}{https://huggingface.co/Jzuluaga/accent-id-commonaccent\_ecapa}}~\cite{zuluaga2023commonaccent}, it is pretrained for English-accent classification, including Singlish~\cite{zuluaga2023commonaccent}. Since the encoder reaches roughly $\sim$90\% accent-classification accuracy and clusters by phonological similarity, cosine distance in its space functions as an accent-fidelity score. The central question of the paper: how closely does generated speech reproduce the Singlish accent? Thus, the choice of ACC-SIM is motivated by~\cite{huang2026codecmos}, where the analogous O-ACC-SIM was shown to track subjective accent-similarity judgments at scale \cite{huang2026codecmos}.

\section{Results}
Table~\ref{tab:total_metrics} reports all four metrics for real speech (Ground Truth), the off-the-shelf base models, and their fine-tuned counterparts, on seen speakers during fine-tuning (in-domain) and unseen speakers (out-of-domain). Similarity is measured against the inference \textit{prompt} and against the held-out target (\textit{match}); Ground Truth attains \textit{match} equal 1 by construction, and its \textit{prompt} values give a same-speaker, cross-content reference baseline.
 
\textbf{Accent and Speaker Similarity.} ACC-SIM (\textit{match}) is our
primary indicator of accent transfer. Off-the-shelf, Chatterbox is weakest
($0.5114$/$0.4697$), confirming that single-shot prompting does not induce a
Singlish accent; fine-tuning yields the largest gain ($+0.126$/$+0.092$, to $0.6376$/$0.5617$). CosyVoice starts higher ($0.5771$/$0.5212$) and gains less ($+0.027$/$+0.028$), consistent with
broader base-model coverage and less headroom. Speaker similarity follows the same pattern (Chatterbox $0.6036\!\rightarrow\!0.6949$ in-domain; CosyVoice nearly flat). On the \textit{prompt} ACC-SIM metric F-Chatterbox even exceeds the Ground Truth reference baselines ($0.6748$ vs. $0.6277$), whereas
F-CosyVoice is \emph{flat} ($0.6200$ vs. $0.6277$, $0.5820$ vs. $0.5802$) while its \textit{match} ACC-SIM holds a drift toward a generalized Singlish voice rather than the specific prompt speaker.

% Reason: the prompt anchor is cross-utterance — real speech is scored against a different genuine clip of the same speaker (real within-speaker variability across content/session lowers it), while zero-shot TTS is conditioned on that exact prompt and regresses toward it, producing output more self-similar to the prompt than an independent natural token is. So "beating GT" is an anchor artifact, not super-human accent transfer; only within-anchor comparisons are valid, and the matched anchor (GT = 1.0) is the true ceiling.

% Because generated speech is conditioned on the prompt, it can inherit prompt-specific acoustic detail that two independent real recordings do not share; the Ground-Truth \textit{prompt} values are therefore a soft same-speaker reference rather than a strict upper bound, and F-Chatterbox exceeding them reflects prompt-consistency, not supra-natural accent.

\textbf{Intelligibility and Naturalness.} The base models fail on WER in
opposite directions. Off-the-shelf Chatterbox is \emph{more} intelligible
than real speech ($11.48\%$ vs.\ $17.31\%$) by producing clean, accent-neutral
output; fine-tuning raises its WER toward the real band ($15.39\%$). CosyVoice instead mis-renders content from the prompt ($19.20\%$). We found CosyVoice is unstable on short utterances and hallucinating, which leads to inflated WER. Fine-tuning \emph{lowers} its WER to $14.68\%$. UT-MOS is reported with caution: real Singlish scores \emph{lowest} ($2.50$/$2.37$)
and the off-the-shelf models highest ($3.33$ - $3.35$), reflecting the
predictor's bias toward clean synthetic audio~\cite{huang2026codecmos}.
Fine-tuning moves both toward those of real Singlish recordings (most for F-Chatterbox,
$3.34\!\rightarrow\!2.75$). However, this does not establish that the generated speech sounds more like real Singlish or is more natural to listeners.
 
\textbf{Adaptation and Consistency.} In-domain scores exceed out-of-domain throughout, Fig. \ref{fig:violin_accent}, but the fine-tuned models retain their advantage on unseen speakers - F-Chatterbox's out-of-domain accent similarity ($0.5617$) still beats off-the-shelf in-domain ($0.5114$). Therefore, the adaptation generalises rather than memorises. Finally, F-CosyVoice-idx, which conditions on a learned per-speaker representation rather than the prompt, attains the best \textit{match} similarity (SPK-SIM $0.7383$, \mbox{ACC-SIM} $0.6565$) and the
lowest fine-tuned WER ($11.92$) but the lowest \textit{prompt} similarity,
taking the drift above to its limit and applying only to seen speakers.

\section{Conclusions}
% Choose 1 conclusion paragraph below:

We presented the first systematic study of Singlish-accented zero-shot TTS. Off-the-shelf Chatterbox and CosyVoice~3, prompted with genuine Singlish references, preserve speaker timbre but flatten the accent - confirmed by low accent similarity against matched real recordings. Fine-tuning on 50 IMDA NSC~\cite{koh2019imda} speakers substantially closes this gap: matched ACC-SIM rises by up to $+0.13$ in-domain, and the gain persists on 42 unseen speakers, showing the models generalize accent beyond the training speakers. None of the two backbones is uniformly superior. They fail and recover differently - Chatterbox trades accent-neutral over-intelligibility for authentic Singlish delivery, while CosyVoice's fine-tuning also repairs its hallucination-driven WER and speaker-index conditioning (F-CosyVoice-idx) pushes accent fidelity furthest at the cost of prompt faithfulness and open-set use.

While accent adaptation improves Singlish voice cloning, it also raises potential impersonation and misuse risks. Responsible deployment and release should therefore consider speaker consent, dataset licensing, and safeguards against unauthorized voice cloning.

% We presented the first systematic study of Singlish-accented zero-shot TTS. Fine-tuning on 50 IMDA NSC speakers successfully adapted both Chatterbox and CosyVoice~3 to the Singlish accent for both in-domain and out-of-domain speakers, demonstrating that the learned accent generalizes beyond the training speakers rather than merely memorizing speaker characteristics. Chatterbox exhibited substantially larger gains in accent similarity (+0.126 in-domain and +0.092 out-of-domain), whereas CosyVoice~3 started from a stronger baseline but achieved more modest improvements (+0.027 on both sets). The two backbones also exhibited different trade-offs: Chatterbox exchanged accent-neutral intelligibility for more authentic Singlish delivery, whereas CosyVoice3 fine-tuning additionally mitigated hallucination-driven WER. Finally, speaker-index conditioning (F-CosyVoice-idx) achieved the highest accent fidelity, albeit at the expense of prompt faithfulness and applicability to unseen speakers.

\section{Limitations}
Our evaluation relies entirely on objective metrics. \mbox{ACC-SIM} serves as a proxy for accent similarity but has not been validated against judgments from Singlish-listeners in this study. Likewise, UT-MOS does not establish subjective naturalness for spontaneous Singlish. A formal subjective listening study remains future work.

\section*{AI-Generated Content Disclosure}
The authors used Anthropic's Claude and OpenAI's \mbox{ChatGPT} as writing and editing assistants in preparing this manuscript. All AI-assisted text and figures were reviewed, verified, and edited by the authors, who take full responsibility for the content.
% The authors used Anthropic's Claude and OpenAI's ChatGPT as writing and editing assistants in preparing this manuscript. The tools were used to draft and revise prose in the abstract, introduction, related work, results, and discussion sections, and to generate and refine the plotting code for the figures from numerical results produced by the authors. All research design, experiments, data, quantitative results, and scientific claims are the authors' own; all AI-assisted text and figures were reviewed, verified, and edited by the authors, who take full responsibility for the content.

% \section*{Acknowledgment}

% The preferred spelling of the word ``acknowledgment'' in America is without 
% an ``e'' after the ``g''. Avoid the stilted expression ``one of us (R. B. 
% G.) thanks $\ldots$''. Instead, try ``R. B. G. thanks$\ldots$''. Put sponsor 
% acknowledgments in the unnumbered footnote on the first page.

\bibliographystyle{IEEEtran_title_case}
\bibliography{IEEEabrv,IEEEbiblio}
\end{document}